\documentclass[conference]{IEEEtran}
\IEEEoverridecommandlockouts
\usepackage{cite}
\usepackage{amsmath,amssymb,amsfonts}
\usepackage{algorithmic}
\usepackage{graphicx}
\usepackage{textcomp}
\usepackage{subcaption}
\usepackage{xcolor}
\usepackage{hyperref}
\def\BibTeX{{\rm B\kern-.05em{\sc i\kern-.025em b}\kern-.08em
    T\kern-.1667em\lower.7ex\hbox{E}\kern-.125emX}}
\begin{document}

\title{EmoAugNet: A Signal-Augmented Hybrid CNN-LSTM Framework for Speech Emotion Recognition \\
\thanks{}
}

\author{\IEEEauthorblockN{1\textsuperscript{st} Durjoy Chandra Paul}
\IEEEauthorblockA{\textit{Department of Computer Science \& Engineering} \\
\textit{Shahjalal University of Science \& Technology}\\
Sylhet, Bangladesh \\
pauldurjoychandra@gmail.com}
\and
\IEEEauthorblockN{2\textsuperscript{nd} Gaurob Saha}
\IEEEauthorblockA{\textit{Department of Computer Science \& Engineering} \\
\textit{Columbus State University	}\\
Georgia, USA \\
saha\_gaurob@students.columbusstate.edu	}
\and
\IEEEauthorblockN{3\textsuperscript{rd} Md Amjad Hossain	}
\IEEEauthorblockA{\textit{dept. name of organization (of Aff.)} \\
\textit{name of organization (of Aff.)}\\
City, Country \\
email address or ORCID}
}
\author{
Durjoy Chandra Paul\textsuperscript{1}, Gaurob Saha\textsuperscript{2},  and Md Amjad Hossain\textsuperscript{3} \\
\textsuperscript{1,2,3}\textit{Department of Computer Science \& Engineering}, \\
\textsuperscript{1}\textit{Shahjalal University of Science \& Technology}, Sylhet, Bangladesh \\
\textsuperscript{2,3}\textit{Columbus State University}, Georgia, USA \\
Email- pauldurjoychandra@gmail.com, saha\_gaurob@students.columbusstate.edu, hossain\_mdamjad@columbusstate.edu	
}

\maketitle

\begin{abstract}
Recognizing emotional signals in speech has a significant impact on enhancing the effectiveness of human-computer interaction (HCI). This study introduces EmoAugNet, a hybrid deep learning framework, that incorporates Long Short-Term Memory (LSTM) layers with one-dimensional Convolutional Neural Networks (1D-CNN) to enable reliable Speech Emotion Recognition (SER). The quality and variety of the features that are taken from speech signals have a significant impact on how well SER systems perform. A comprehensive speech data augmentation strategy was used to combine both traditional methods, such as noise addition, pitch shifting, and time stretching, with a novel combination-based augmentation pipeline to enhance generalization and reduce overfitting. Each audio sample was transformed into a high-dimensional feature vector using root mean square energy (RMSE), Mel-frequency Cepstral Coefficient (MFCC), and zero-crossing rate (ZCR). Our model with ReLU activation has a weighted accuracy of 95.78\% and unweighted accuracy of 92.52\% on the IEMOCAP dataset and, with ELU activation, has a weighted accuracy of 96.75\% and unweighted accuracy of 91.28\%. On the RAVDESS dataset, we get a weighted accuracy of 94.53\% and 94.98\% unweighted accuracy for ReLU activation and 93.72\% weighted accuracy and 94.64\% unweighted accuracy for ELU activation. These results highlight EmoAugNet's effectiveness in improving the robustness and performance of SER systems through integated data augmentation and hybrid modeling.
\end{abstract}


\section{Introduction}
Emotion detection refers to the task of identifying a person’s emotional state, such as anger, fear, neutrality, happiness, disgust, or sadness \cite{mande2019emotion}. Speech Emotion Recognition (SER), which records and interprets speech, is essential for enhancing human-computer interaction \cite{el2011survey}. Several applications—like intelligent robotics, audio monitoring, law enforcement, smart home control, and content recommendation systems—depend on recognizing the user's emotions through speech \cite{ahmed2023ensemble}. In the past few years, deep learning has investigated tremendous advances in speech recognition \cite{amodei2016deep,medennikov2016improving,saon2015ibm,liptchinsky2017based}. However, compared to general speech recognition, there are fewer large-scale datasets available for speech emotion recognition, which makes data augmentation an important research topic to evaluate its effect on SER performance \cite{atmaja2022effects}. In the field of SER, EmoAugNet focused on addressing the data shortage problem  through integated data augmentation and enhancing the robustness and performance of SER systems by hybrid modeling. We conducted our experiments on two widely used emotional speech datasets: RAVDESS\cite{livingstone2018ryerson} and IEMOCAP\cite{busso2008iemocap} to evaluate the efficiency of our proposed model.

Key contributions of this research include:

\begin{itemize}
    \item  Introduction of a hybrid deep learning framework for efficient SER that combines LSTM networks with one-dimensional Convolutional Neural Networks (1D-CNN).
    \item Development of a comprehensive combination-based speech data augmentation pipeline consisting of noise injection with pitch shifting as well as both slow and fast time stretching and signal shifting address the training data shortage problem.
    \item  Investigation into the impact of activation functions on SER performance, with observations highlighting their varying effectiveness across different datasets.
\end{itemize}
 To extract features, we utilized ZCR, RMSE, and MFCC to measure temporal characteristics, energy distribution and spectral patterns of speech signals. To train the proposed model, each original and augmented sample was converted into a high-dimensional feature vector. For the purpose of this study, we focused on classifying seven distinct emotions that are consistently present across both datasets: neutral, surprise, disgust, fear, sad, happy, and angry.
\begin{figure*}[h]
  \centering
  \includegraphics[width=0.8\linewidth, height=6cm]{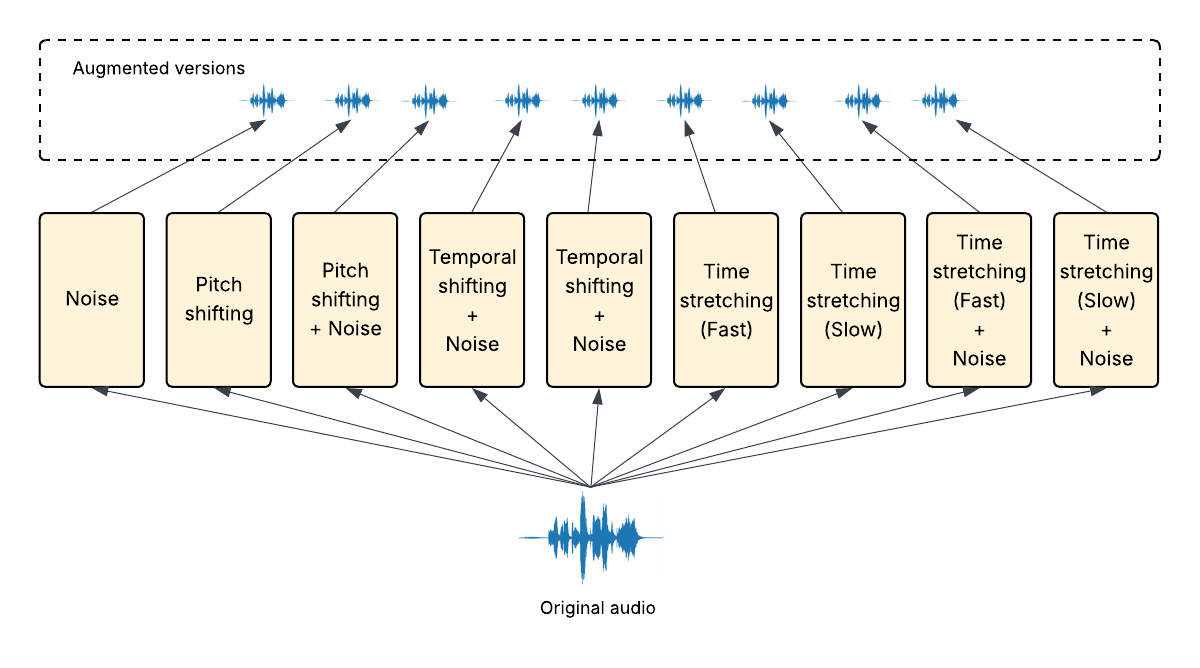}
  \caption{Data Augmentation Pipeline}
  \label{fig:se_model}
\end{figure*}
\section{RELATED WORKS}
In the existing literature, a significant number of researchers have developed speech-based emotion recognition systems using  CNNs, RNNs, DNNs, and LSTMs\cite{ma2018emotion}. In addition to these standalone approaches, some studies have investigated the effectiveness of combining multiple architectures to capture both spatial and temporal features more comprehensively \cite{sajjad2020clustering} \cite{liu2018eera}.
 Etienne et al. (2018) suggested a hybrid neural model that integrates convolutional layers to extract spectral features and LSTM layers to capture temporal relationships in speech signals. The study used the IEMOCAP dataset and performed layer-wise optimization adjustments, batch normalization in the recurrent layers, and vocaltract length perturbation for data augmentation. The model achieved 61.7\% unweighted accuracy for four emotion classes, combining CNN and LSTM layers with tailored training techniques for improving performance in speech emotion recognition \cite{etienne2018cnn+}. Mustaqeem and Kwon proposed a ConvLSTM-based SER model with hierarchical local feature learning blocks (LFLBs) to capture spatiotemporal emotional cues. They used residual learning, sequence modeling, and combined center loss with softmax to enhance classification. Their model achieved 75\% and 80\% accuracy on IEMOCAP and RAVDESS, respectively \cite{mustaqeem2020clstm}. The majority of the researchers believe that continuous speech characteristics, such as energy and pitch, aid in conveying the feelings that underlies what is being said \cite{cowie2001emotion,busso2009analysis,ten2003emotions}. Speech emotion recognition has made extensive use of continuous speech features. For instance, Banse et al. investigated vocal indicators across 14 emotion categories \cite{banse1996acoustic}. Fundamental frequency (F0), energy levels, articulation rate, and spectral properties were among the attributes they concentrated on. Anusha Koduru et al. extracted a comprehensive set of speech features—including MFCC, pitch, energy, ZCR, and DWT —to capture emotional characteristics. Their approach achieved up to 85\% accuracy on the RAVDESS dataset using DTree\cite{koduru2020feature}. For data augmentation in Speech Emotion Recognition (SER), Atmaja et al. (2022) applied several techniques, including glottal source extraction, speech cleaning, impulse response, and noise addition, all combined with the original IEMOCAP dataset. The highest unweighted average recall (UAR) of 75.87\% was obtained when the original data was augmented using a combination of speech cleaning, impulse response, and noise addition \cite{atmaja2022effects}.

\section{Data augmentation}
Both RAVDESS and IEMOCAP are relatively small datasets, which poses a challenge when training neural network models. To address this limitation, data augmentation techniques are commonly employed to compensate for the limited data. In this work, the augmentation process began with the addition of Gaussian noise to the original waveform. The noise was scaled using a constant factor of 0.035 to keep the balance between distortion and realism. It was then multiplied by a random number between 0 and 1 and by the peak amplitude of the signal. 
\begin{itemize}
\item Pitch shifting was applied using a pitch factor randomly selected from the range [-1, 1], allowing both upward and downward shifts of the original frequency components, thus simulating natural variations in speaker tone.

\item  Time stretching was also implemented to simulate changes in speech tempo: the audio was either slowed down using a stretch rate of 0.9 or sped up using a rate of 1.1, as commonly recommended in prior works \cite{jaitly2013vocal ,cui2015data}.
\item Additionally, random temporal shifting was performed by rolling the waveform forward or backward by up to ±5000 samples, introducing timing variability similar to misalignments or speech onset variations in real-world recordings.

\end{itemize}

  These augmentations were also combined in different ways, such as adding noise to audio that was already pitch-shifted or time-stretched. After applying these augmentations and their combinations, a total of 10 audio samples were created from each original audio file, ( Fig. 1).

\begin{figure}[h]
  \centering
  \includegraphics[width=\linewidth, height=2cm]{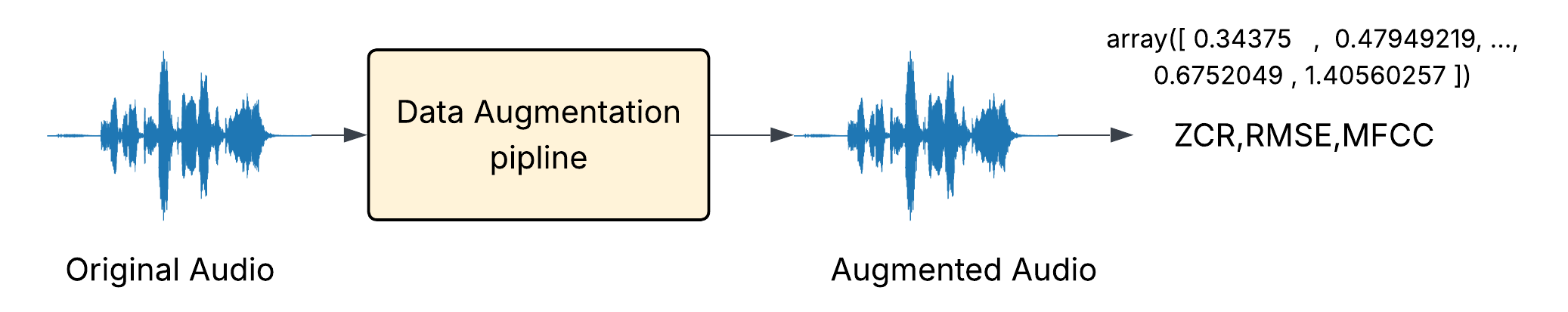}
  \caption{Feature Extraction }
  \label{fig:feature_extraction}
\end{figure}

\begin{figure*}[h]
  \centering
  \includegraphics[width=0.9 \linewidth, height=5cm]{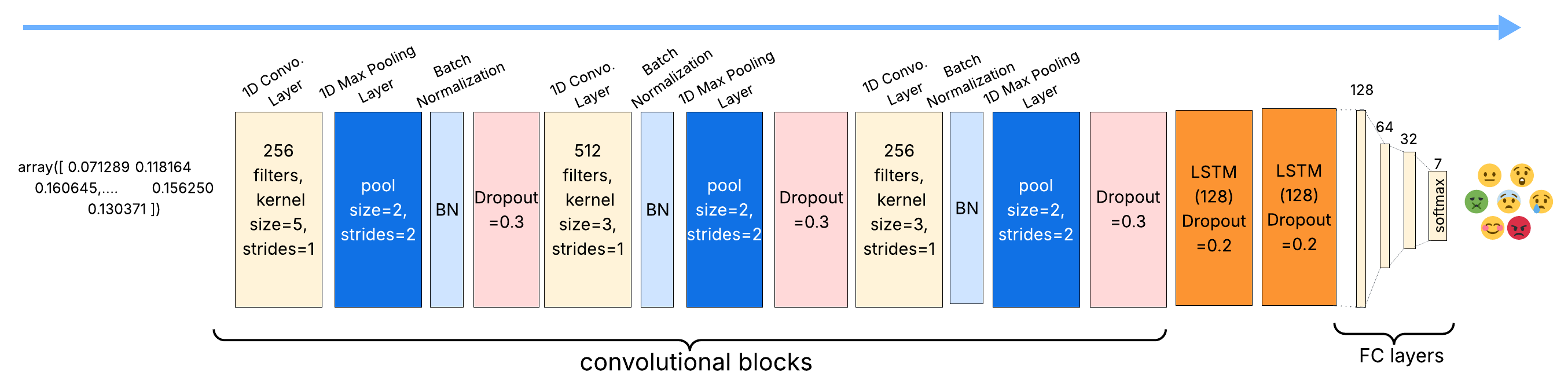}
  \caption{Architecture of Conv1D-LSTM  model}
  \label{fig:se_model}
\end{figure*}
\section{Model description}

\begin{table*}[ht]
\centering
\caption{Conv1D-LSTM Architecture Summary}
\begin{tabular}{|l|c|c|}
\hline
\textbf{Layer (type)} & \textbf{Output Shape} & \textbf{Parameters} \\
\hline
Conv1D & (None, 2376, 256) & 1,536 \\
MaxPooling1D & (None, 1188, 256) & 0 \\
BatchNormalization & (None, 1188, 256) & 1,024 \\
Dropout & (None, 1188, 256) & 0 \\
Conv1D & (None, 1188, 512) & 393,728 \\
BatchNormalization & (None, 1188, 512) & 2,048 \\
MaxPooling1D & (None, 594, 512) & 0 \\
Dropout & (None, 594, 512) & 0 \\
Conv1D & (None, 594, 256) & 393,472 \\
BatchNormalization & (None, 594, 256) & 1,024 \\
MaxPooling1D & (None, 297, 256) & 0 \\
LSTM & (None, 297, 128) & 197,120 \\
Dropout & (None, 297, 128) & 0 \\
LSTM & (None, 128) & 131,584 \\
Dropout & (None, 128) & 0 \\
Dense & (None, 128) & 16,512 \\
BatchNormalization & (None, 128) & 512 \\
Dense & (None, 64) & 8,256 \\
BatchNormalization & (None, 64) & 256 \\
Dense & (None, 32) & 2,080 \\
BatchNormalization & (None, 32) & 128 \\
Dropout & (None, 32) & 0 \\
Dense & (None, 7) & 231 \\
\hline
\end{tabular}
\label{table:model_summary}
\end{table*}
The architecture is designed for effective Speech Emotion Recognition (SER) by combining convolutional and recurrent layers to extract both short-term acoustic patterns and long-term temporal dependencies from speech. After each audio augmentation, features such as the ZCR, RMSE, and MFCC were extracted. These features were computed frame-by-frame to represent the signal’s energy, frequency content, and spectral shape, which are essential for distinguishing different emotions ( Fig. 2). ZCR helps to capture the noise and frequency characteristics of speech, RMSE reflects signal energy, and MFCC encodes spectral features relevant to emotional tone. These features were stacked sequentially to maintain the temporal structure of the speech signal.

To learn localized features from these sequences, a series of 1D convolutional layers was used. Smaller kernel sizes (5 and 3) with a stride of 1 allow the model to focus on fine-grained changes in the signal, such as pitch and energy variations. In terms of dimensionality reduction, a max-pooling 
layer was employed following each CNN layer, with a stride of 2, to retain the most important features of the data. Batch normalization was added after each convolution to stabilize learning and dropout was applied to prevent overfitting.

As part of architectural experimentation, different activation functions were applied within the convolutional blocks to observe their impact on the model performance. It was found that our model achieved  higher accuracy on the RAVDESS dataset when ReLU activation functions were used in the convolutional layers, likely due to the dataset's controlled and high-quality recordings. Conversely, we got improved results on the IEMOCAP dataset when ELU activation functions were employed, the credit can be attributed to ELU’s capacity to better handle more expressive, spontaneous, and noisy speech samples. 

After the convolutional blocks, the input sequence's temporal progression of emotional patterns is captured by two LSTM layers. This helps the model retain and use the information from earlier and later parts of the audio signal. The output is then passed through a series of fully connected (FC) layers with ELU activation functions, which show consistent improvement (around 2–4\%)in recognition accuracy compared to the ReLU activation functions' recognition accuracy when tested across multiple datasets. These FC layers help to refine the learned representations before the final softmax layer, which performs multi-class classification into seven emotion categories( Fig. 3).

\section{EXPERIMENTAL RESULTS AND DISCUSSION}
\begin{table*}[ht]
\centering
\caption{Comparison of Methods on IEMOCAP and RAVDESS Datasets}
\begin{tabular}{|c|l|c|c|c|}
\hline
\textbf{Dataset} & \textbf{Method} & \textbf{Classes} & \textbf{Weighted Accuracy} & \textbf{Unweighted Accuracy} \\
\hline
IEMOCAP & CNN--LSTM~\cite{satt2017efficient} & 6 & 51.3\% & 46.5\% \\
IEMOCAP & DST~\cite{chen2023dst} & 4 & 71.2\% & 72.9\% \\
IEMOCAP & MFGCN~\cite{qi2025mfgcn} & 6 & 65.5\% & 65.8\% \\
IEMOCAP & Proposed (with ReLU) & 7 & \textbf{95.78\%} & \textbf{92.52\%} \\
IEMOCAP & Proposed (with ELU) & 7 & \textbf{96.75\%} & \textbf{91.28\%} \\
RAVDESS & CNN--LSTM~\cite{satt2017efficient} & 4 & 63.2\% & 60.7\% \\
RAVDESS & MFGCN~\cite{qi2025mfgcn} & 4 & 85.7\% & 85.1\% \\
RAVDESS & Proposed (with ReLU) & 7 & \textbf{94.53\%} & \textbf{94.98\%} \\
RAVDESS & Proposed (with ELU) & 7 & \textbf{93.72\%} & \textbf{94.64\%} \\
\hline
\end{tabular}
\label{tab:comparison}
\end{table*}

\begin{figure}[h]
  \centering

  \begin{subfigure}[b]{0.45\textwidth}
    \includegraphics[width=\linewidth]{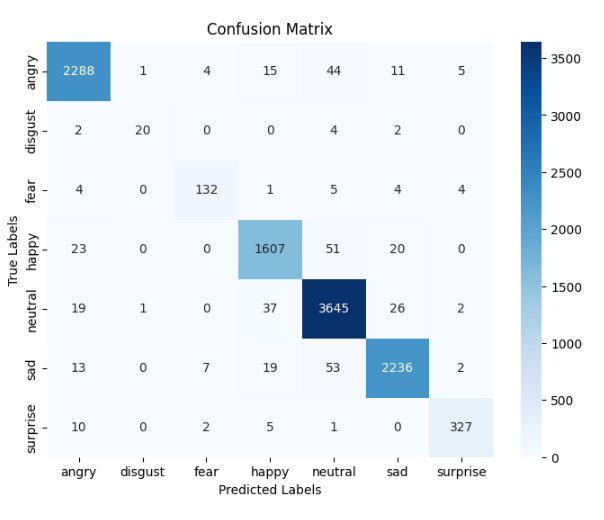}
    \caption{ELU}
  \end{subfigure}
  \hfill
  \begin{subfigure}[b]{0.45\textwidth}
    \includegraphics[width=\linewidth]{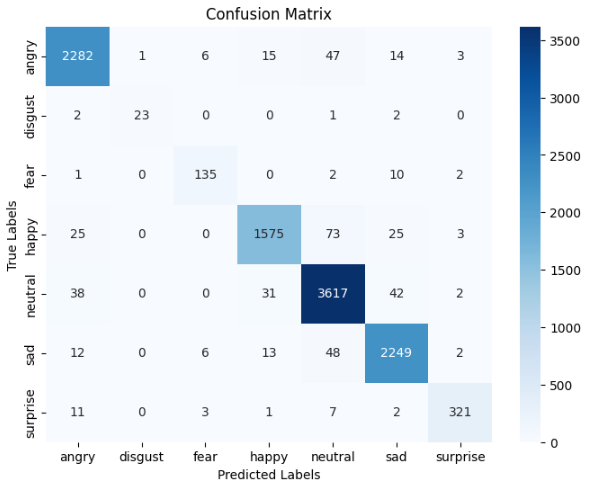}
    \caption{ReLU}
  \end{subfigure}

  \caption{Comparison of confusion matrices on the \textbf{IEMOCAP} dataset, using different activation functions.}
  \label{fig:confusion_matrices}
\end{figure}

\begin{figure}[h]
  \centering

  \begin{subfigure}[b]{0.45\textwidth}
    \includegraphics[width=\linewidth]{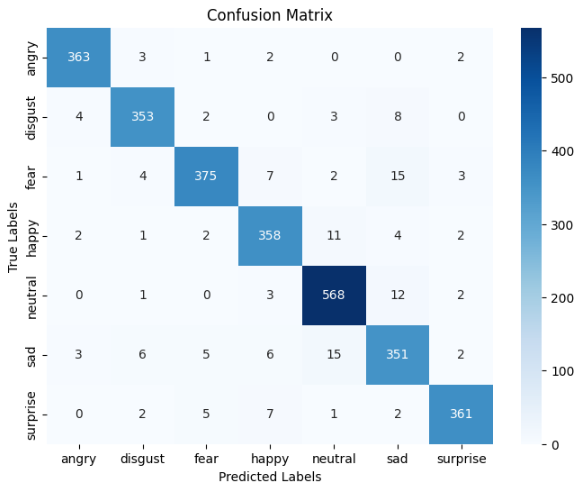}
    \caption{ELU}
  \end{subfigure}
  \hfill
  \begin{subfigure}[b]{0.45\textwidth}
    \includegraphics[width=\linewidth]{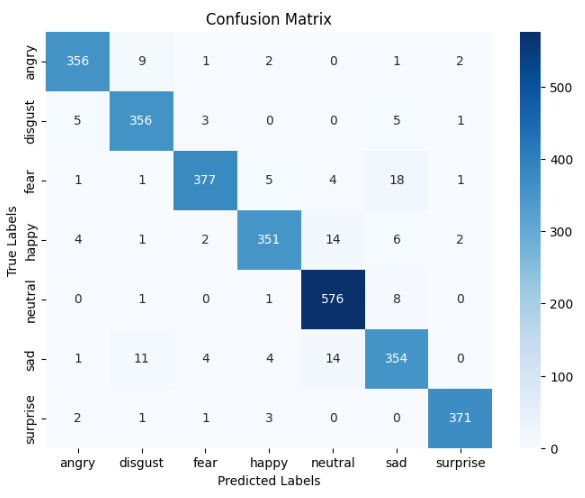}
    \caption{ReLU}
  \end{subfigure}

  \caption{Comparison of confusion matrices on the \textbf{RAVDESS} dataset, using different activation functions.}
  \label{fig:confusion_matrices}
\end{figure}
We evaluated our proposed Conv1D-LSTM model on the RAVDESS and IEMOCAP datasets by varying different activation functions in the Conv1D layers and incorporating different augmentation techniques. Training was done using a learning rate of 0.001. To ensure stable and efficient convergence, the learning rate was automatically reduced when the validation accuracy stopped improving. Early stopping was also employed to avoid overfitting and maintain the top-performing model weights throughout the training process. Using the RAVDESS dataset, the model obtained a weighted accuracy of 93.72\% and unweighted accuracy of 94.64\% using ELU activation function, while replacing ELU with ReLU slightly improved the performance to a weighted accuracy of 94.53\% and unweighted accuracy of 94.98\%. On the IEMOCAP dataset, the model attained a weighted accuracy of 96.75\% and unweighted accuracy of 91.28\% with ELU, whereas using ReLU resulted in 95.78\% weighted and 92.52\% unweighted accuracy. We compared the performance of our Conv1D-LSTM model with a number of cutting-edge methods from the literature in order to further confirm its efficacy ( Table II).

\section{Conclusion}
 We present a hybrid Conv1D-LSTM architecture for SER in this work, enhanced by a novel data augmentation pipeline combining noise injection, pitch shifting, time stretching, and temporal shifting. Our approach leverages ZCR, RMSE, and MFCC features to capture spectral and temporal emotion cues, achieving state-of-the-art performance on both IEMOCAP and RAVDESS datasets. By integrating convolutional layers for local feature extraction and LSTM layers for sequential modeling, along with regularization techniques such as batch normalization and dropout, our architecture effectively addresses the challenges of variability in speech signals. In the future, our goal will be to explore multilingual and cross-lingual emotion recognition to assess the adaptability of the model across languages. Furthermore, our objective is to enhance the robustness and practical utility of this model by incorporating more diverse and comprehensive datasets.

\bibliographystyle{IEEEtran}
\bibliography{custom}

\begin{thebibliography}{10}
\providecommand{\url}[1]{#1}
\csname url@samestyle\endcsname
\providecommand{\newblock}{\relax}
\providecommand{\bibinfo}[2]{#2}
\providecommand{\BIBentrySTDinterwordspacing}{\spaceskip=0pt\relax}
\providecommand{\BIBentryALTinterwordstretchfactor}{4}
\providecommand{\BIBentryALTinterwordspacing}{\spaceskip=\fontdimen2\font plus
\BIBentryALTinterwordstretchfactor\fontdimen3\font minus \fontdimen4\font\relax}
\providecommand{\BIBforeignlanguage}[2]{{%
\expandafter\ifx\csname l@#1\endcsname\relax
\typeout{** WARNING: IEEEtran.bst: No hyphenation pattern has been}%
\typeout{** loaded for the language `#1'. Using the pattern for}%
\typeout{** the default language instead.}%
\else
\language=\csname l@#1\endcsname
\fi
#2}}
\providecommand{\BIBdecl}{\relax}
\BIBdecl

\bibitem{mande2019emotion}
A.~A. Mande, S.~Dani, S.~Telang \emph{et~al.}, ``Emotion detection using audio data samples.'' \emph{International journal of advanced research in computer science}, vol.~10, no.~6, 2019.

\bibitem{el2011survey}
M.~El~Ayadi, M.~S. Kamel, and F.~Karray, ``Survey on speech emotion recognition: Features, classification schemes, and databases,'' \emph{Pattern recognition}, vol.~44, no.~3, pp. 572--587, 2011.

\bibitem{ahmed2023ensemble}
M.~R. Ahmed, S.~Islam, A.~M. Islam, and S.~Shatabda, ``An ensemble 1d-cnn-lstm-gru model with data augmentation for speech emotion recognition,'' \emph{Expert Systems with Applications}, vol. 218, p. 119633, 2023.

\bibitem{amodei2016deep}
D.~Amodei, S.~Ananthanarayanan, R.~Anubhai, J.~Bai, E.~Battenberg, C.~Case, J.~Casper, B.~Catanzaro, Q.~Cheng, G.~Chen \emph{et~al.}, ``Deep speech 2: End-to-end speech recognition in english and mandarin,'' in \emph{International conference on machine learning}.\hskip 1em plus 0.5em minus 0.4em\relax PMLR, 2016, pp. 173--182.

\bibitem{medennikov2016improving}
I.~Medennikov, A.~Prudnikov, and A.~Zatvornitskiy, ``Improving english conversational telephone speech recognition.'' in \emph{Interspeech}, 2016, pp. 2--6.

\bibitem{saon2015ibm}
G.~Saon, H.-K.~J. Kuo, S.~Rennie, and M.~Picheny, ``The ibm 2015 english conversational telephone speech recognition system,'' \emph{arXiv preprint arXiv:1505.05899}, 2015.

\bibitem{liptchinsky2017based}
V.~Liptchinsky, G.~Synnaeve, and R.~Collobert, ``based speech recognition with gated convnets,'' \emph{arXiv preprint arXiv:1712.09444}, 2017.

\bibitem{atmaja2022effects}
B.~T. Atmaja and A.~Sasou, ``Effects of data augmentations on speech emotion recognition,'' \emph{Sensors}, vol.~22, no.~16, p. 5941, 2022.

\bibitem{livingstone2018ryerson}
S.~R. Livingstone and F.~A. Russo, ``The ryerson audio-visual database of emotional speech and song (ravdess): A dynamic, multimodal set of facial and vocal expressions in north american english,'' \emph{PloS one}, vol.~13, no.~5, p. e0196391, 2018.

\bibitem{busso2008iemocap}
C.~Busso, M.~Bulut, C.-C. Lee, A.~Kazemzadeh, E.~Mower, S.~Kim, J.~N. Chang, S.~Lee, and S.~S. Narayanan, ``Iemocap: Interactive emotional dyadic motion capture database,'' \emph{Language resources and evaluation}, vol.~42, pp. 335--359, 2008.

\bibitem{ma2018emotion}
X.~Ma, Z.~Wu, J.~Jia, M.~Xu, H.~Meng, and L.~Cai, ``Emotion recognition from variable-length speech segments using deep learning on spectrograms.'' in \emph{Interspeech}, 2018, pp. 3683--3687.

\bibitem{sajjad2020clustering}
M.~Sajjad, S.~Kwon \emph{et~al.}, ``Clustering-based speech emotion recognition by incorporating learned features and deep bilstm,'' \emph{IEEE access}, vol.~8, pp. 79\,861--79\,875, 2020.

\bibitem{liu2018eera}
B.~Liu, H.~Qin, Y.~Gong, W.~Ge, M.~Xia, and L.~Shi, ``Eera-asr: An energy-efficient reconfigurable architecture for automatic speech recognition with hybrid dnn and approximate computing,'' \emph{IEEE Access}, vol.~6, pp. 52\,227--52\,237, 2018.

\bibitem{etienne2018cnn+}
C.~Etienne, G.~Fidanza, A.~Petrovskii, L.~Devillers, and B.~Schmauch, ``Cnn+ lstm architecture for speech emotion recognition with data augmentation,'' \emph{arXiv preprint arXiv:1802.05630}, 2018.

\bibitem{mustaqeem2020clstm}
Mustaqeem and S.~Kwon, ``Clstm: Deep feature-based speech emotion recognition using the hierarchical convlstm network,'' \emph{Mathematics}, vol.~8, no.~12, p. 2133, 2020.

\bibitem{cowie2001emotion}
R.~Cowie, E.~Douglas-Cowie, N.~Tsapatsoulis, G.~Votsis, S.~Kollias, W.~Fellenz, and J.~G. Taylor, ``Emotion recognition in human-computer interaction,'' \emph{IEEE Signal processing magazine}, vol.~18, no.~1, pp. 32--80, 2001.

\bibitem{busso2009analysis}
C.~Busso, S.~Lee, and S.~Narayanan, ``Analysis of emotionally salient aspects of fundamental frequency for emotion detection,'' \emph{IEEE transactions on audio, speech, and language processing}, vol.~17, no.~4, pp. 582--596, 2009.

\bibitem{ten2003emotions}
L.~Ten~Bosch, ``Emotions, speech and the asr framework,'' \emph{Speech Communication}, vol.~40, no. 1-2, pp. 213--225, 2003.

\bibitem{banse1996acoustic}
R.~Banse and K.~R. Scherer, ``Acoustic profiles in vocal emotion expression.'' \emph{Journal of personality and social psychology}, vol.~70, no.~3, p. 614, 1996.

\bibitem{koduru2020feature}
A.~Koduru, H.~B. Valiveti, and A.~K. Budati, ``Feature extraction algorithms to improve the speech emotion recognition rate,'' \emph{International Journal of Speech Technology}, vol.~23, no.~1, pp. 45--55, 2020.

\bibitem{jaitly2013vocal}
N.~Jaitly and G.~E. Hinton, ``Vocal tract length perturbation (vtlp) improves speech recognition,'' in \emph{Proc. ICML workshop on deep learning for audio, speech and language}, vol. 117, 2013, p.~21.

\bibitem{cui2015data}
X.~Cui, V.~Goel, and B.~Kingsbury, ``Data augmentation for deep neural network acoustic modeling,'' \emph{IEEE/ACM Transactions on Audio, Speech, and Language Processing}, vol.~23, no.~9, pp. 1469--1477, 2015.

\bibitem{satt2017efficient}
A.~Satt, S.~Rozenberg, R.~Hoory \emph{et~al.}, ``Efficient emotion recognition from speech using deep learning on spectrograms.'' in \emph{Interspeech}, 2017, pp. 1089--1093.

\bibitem{chen2023dst}
W.~Chen, X.~Xing, X.~Xu, J.~Pang, and L.~Du, ``Dst: Deformable speech transformer for emotion recognition,'' in \emph{ICASSP 2023-2023 IEEE International Conference on Acoustics, Speech and Signal Processing (ICASSP)}.\hskip 1em plus 0.5em minus 0.4em\relax IEEE, 2023, pp. 1--5.

\bibitem{qi2025mfgcn}
X.~Qi, Y.~Wen, P.~Zhang, and H.~Huang, ``Mfgcn: Multimodal fusion graph convolutional network for speech emotion recognition,'' \emph{Neurocomputing}, vol. 611, p. 128646, 2025.

\end{thebibliography}

\end{document}